\documentclass[sigconf,screen]{acmart} 

\usepackage{hyperref}
\hypersetup{breaklinks=true}

\usepackage{balance}
\usepackage{amsmath,amsfonts}
\usepackage{algorithmic}
\usepackage{graphicx}
\usepackage{textcomp}
\usepackage{xcolor}
\def\BibTeX{{\rm B\kern-.05em{\sc i\kern-.025em b}\kern-.08em
    T\kern-.1667em\lower.7ex\hbox{E}\kern-.125emX}}

\usepackage{booktabs}
\usepackage{xspace}
\usepackage{subcaption}
\newcommand{\thistool}{{AdaptivePaste\xspace}}

\title{AdaptivePaste: Code Adaptation through Learning Semantics-aware Variable Usage Representations }

\author{Xiaoyu Liu}
\affiliation{%
  \institution{Microsoft}
  \city{Redmond}
  \state{WA}
  \country{USA}
}
\author{Jinu Jang}
\affiliation{%
  \institution{Microsoft}
  \city{Redmond}
  \state{WA}
  \country{USA}
}
\author{Neel Sundaresan}
\affiliation{%
  \institution{Microsoft}
  \city{Redmond}
  \state{WA}
  \country{USA}
}
\author{Miltiadis Allamanis}
\affiliation{%
  \institution{Google}
  \city{Cambridge}
  \country{UK}
}
\author{Alexey Svyatkovskiy}
\affiliation{%
  \institution{Microsoft}
  \city{Redmond}
  \state{WA}
  \country{USA}
}

\begin{document}

\begin{abstract}
  In software development, it is common for programmers to copy-paste or port code snippets and then adapt them to their use case. This scenario motivates the \textit{code adaptation} task -- a variant of program repair which aims to adapt variable identifiers in a pasted snippet of code to the surrounding, preexisting context. However, no existing approach has been shown to effectively address this task. In this paper, we introduce \thistool, a learning-based approach to source code adaptation, based on transformers and a dedicated dataflow-aware deobfuscation pre-training task to learn meaningful representations of variable usage patterns. We demonstrate that \thistool{} can learn to adapt Python source code snippets with 67.8\% exact match accuracy. We study the impact of confidence thresholds on the model predictions, showing the model precision can be further improved to 85.9\% with our parallel-decoder transformer model in a selective code adaptation setting.
  To assess the practical use of \thistool{} we perform a user study among Python software developers on real-world copy-paste instances. The results show that \thistool{} reduces dwell time to nearly half the time it takes to port code manually, and helps to avoid bugs. In addition, we utilize the participant feedback to identify potential avenues for improvement.
\end{abstract}



\maketitle

\section{Introduction}

Knowledge reuse through adapting source code snippets from Stack Overflow posts or porting code changes from peer software projects on GitHub is very common among software developers. Indeed, according to a study~\cite{stackoverflow_survey}, one out of every four users visiting a Stack Overflow question copies something within five minutes of hitting the page. That adds up to 2.9 million copies across half a million posts and comments on average, every day. An empirical study of GitHub commit histories documented in~\cite{ray12} shows that 11\% to 16\% of all code changes in large software repositories are ported from peer projects. While standing on the shoulders of giants and reusing code from the web or peer projects, developers may still introduce compilation issues, bugs, or even security issues into their code~\cite{ray2013detecting}. One common code adaptation scenario is to manually rename variables in a copied snippet to align them with the surrounding context. Automatically adapting copy-pasted code using AI could reduce errors and relieve developers from the effort required.

The goal of the code adaptation task is to intelligently adjust a given (pasted) snippet to preexisting partially written program~\cite{allamanis2017smartpaste} and has a deep connection to the program repair task. Adapting source code manually is a tedious and error-prone task. Indeed, a software developer must find program locations where buggy variables are used and change them to the correct variables.  There are numerous systems~\cite{allamanis2018learning, allamanis2021self,vasic2019neural,hellendoorn2019global} proposed to tackle variable misuse problem with machine learning. These approaches focus on classifying a single variable location as faulty --- if any --- and then repairing the faulty variables through a joint prediction of classification, localization, and repair~\cite{vasic2019neural}, or an enumerative prediction of each buggy location~\cite{allamanis2018learning}. Unfortunately, these approaches look for a single misused variable and are not well-suited for adapting code snippets that will commonly have multiple misused variables.

In this paper, we propose \thistool: a code adaptation system based on the transformer neural network that adjusts copy-pasted or ported source code snippets to the surrounding code context. Given a partially written program, a code snippet copied from an external source, and a location where the snippet is pasted to, \thistool{} analyzes the code, anonymizes the variables in the pasted code snippet in a dataflow-sensitive way, and learns to predict a mapping of anonymized (masked) variables to the target names via self-supervised translation. Inspired by recent success of transformers~\cite{guo2022learning, vaswani2017attention}, we implement two variants of \thistool: a traditional encoder-decoder transformer to decode all adapted identifiers appearing in a pasted code snippet as one sequence (AP-uni), and a parallel-decoder transformer model to adapt each individual identifier independently (AP-parallel). Unlike traditional encoder-decoder implementation, our parallel decoder (AP-par) variant can recover from providing incorrect identifier adaptations early in decoding process, thus effectively improving code adaptation quality in real-world scenarios. It also allows to achieve the highest code adaptation precision given thresholding model predictions by confidence scores. 

Our contributions are as follows: (i) we formulate a learning-based approach to source code adaptation and introduce \thistool, a transformer-based model of two variants (traditional uni-decoder and parallel-decoder model with tied weights), (ii) we introduce a specialized dataflow-aware deobfuscation pre-training objective, a variant of DOBF~\cite{lachaux2021dobf}, for pasted code snippet adaptation, (iii) we evaluate \thistool{} on a dataset of code snippets in Python and demonstrate that our model achieves as much as 67.8\% accuracy of adapting source code and 83.9\% accuracy of predicting a valid code snippet, both beating existing state-of-the-art models, and (iv) leveraging our parallel-decoder implementation enables \thistool{} to adapt code identifiers selectively, which is especially beneficial when the previous identifiers are predicted incorrectly, and (v) we perform a user study among professional software developers on real-world code snippets which shows practical value of the tool capable of reducing the dwell time spent adapting source code, and reducing errors.

\section{Motivating Example}

To provide the intuition about how our code adaptation approach works and to define the core concepts we begin with a motivating example of a code copy-paste scenario. 

In the code adaptation problem, a pasted code snippet is a piece of code copied from a website (e.g. StackOverflow), a code search retrieval tool, or some other source, which is then inserted into some preexisting context. In a pasted code snippet, we distinguish two kinds of variables to be adapted: (i) \textit{bound variables} -- symbols used or defined in the context falling in the scope of the pasted snippet, and (ii) \textit{free variables} -- identifiers that are defined and used exclusively inside the pasted code snippet.

The context snippet contains a number of variables/symbols $V=\{v_i\}$. The pasted snippet uses and defines a set of
variables $B=\{\beta_j\}$. The \thistool{} task is to create an injective mapping $\mathcal{A}: B\rightarrow V\cup \Sigma$ that maps each variable in the pasted
snippet to either a unique symbol $v_i$ in the context or a declares this as new variable $\Sigma$, not currently defined\footnote{We assume that $\Sigma$ contains all valid variable names \emph{not} in the context.}. Every $\mathcal{A}(\beta_j)\in V$ is a variable \emph{bound} to the context, whereas if $\mathcal{A}(\beta_j)\in \Sigma$ it is a free variable. Note, that for semantic correctness $\mathcal{A}$ must only correctly map the bound variables.
\thistool{} can be seen as the task of correctly ``wiring'' the pasted snippet into the context.
\begin{figure*}
\centering
 \includegraphics[width=0.7\textwidth]{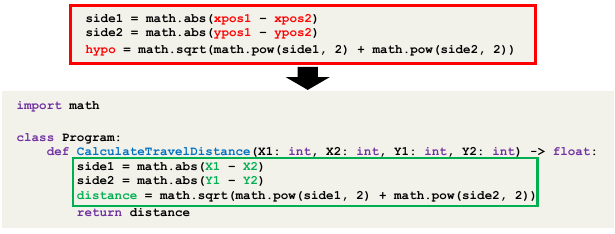}
 \caption{
 An illustration of a code copy-paste scenario when developing software.}
 \label{fig:motivation}
\end{figure*}

\autoref{fig:motivation} illustrates a code copy-paste scenario where a developer first writes a signature of a method named $\texttt{CalculateTravelDistance}$ taking 4 integer parameters -- $X1$, $X2$, $Y1$, and $Y2$, and returning a variable $\texttt{distance}$ of type float. These integer parameters serve as coordinates of the start and end positions of a moving object in a two-dimensional plane, and the return variable corresponds to the distance traveled. A developer then copy-pastes from an external source a code snippet that calculates the Euclidean distance traveled by a moving object (copied region is enclosed in red square box). \thistool{} ``re-wires'' the pasted code snippet to form 
a correct program given the preexisting context -- method signature and return statement -- by adjusting names of certain variables (source variable names are highlighted in red, target variable names are highlighted in green). Namely, variables $\texttt{xpos1}$,  $\texttt{xpos2}$,  $\texttt{ypos1}$,  and $\texttt{ypos2}$ representing the coordinates of the object in two-dimensional plane are tied to the preexisting method arguments, and the variable $\texttt{hypo}$ that stores the resulting distance is bound to the return variable named  $\texttt{distance}$.

\section{Dataset}
We collect a dataset from publicly available open-source non-fork GitHub repositories in Python programming language having licenses explicitly permitting the re-distribution of parts of the project. To ensure high quality of the dataset, we require each project to have at least 20 starts, and to be used by at least one other project. These quality criteria follow closely the definition of~\cite{feng2020codebert}  and \cite{lu2021codexglue}. In addition, we filter out all unit test files and short source code files having less than three lines of code. Overall, we collect $4.3\times10^6$ files having $1.4\times10^9$ lines of code across $238\times10^3$ repositories. For model training and evaluation, we randomly split the dataset on a repository level into training, validation, and test sets in the 80-10-10 proportion. The validation fold is utilized for model selection.

The training and validation samples are extracted in a self-supervised manner. To mimic copy-pasting, we extract samples consisting of a context snippet and a ``pastable’’ block. We parse each source code file in preselected projects and randomly pick one or more contiguous statements of the following non-terminal node types of the \texttt{LibCST}\footnote{\url{https://github.com/Instagram/LibCST}} grammar: \texttt{Module}, \texttt{SimpleStatementSuite}, \texttt{SimpleStatementLine},  \texttt{IndentedBlock}, \texttt{If}, \texttt{With}, \texttt{For}, \texttt{While}, \texttt{Else}, \texttt{Try}, \texttt{Finally},  \texttt{ExceptHandler}. We set the max number of lines of code in a pastable block to 60. In summary, we extract 29,176,497 pastable snippet samples for model training and 1,774,995 samples for model evaluation.

\section{Baseline Models}

We now discuss three baselines that we implemented for controlled experiments around code adaptation. 

\subsection{Masked Language Modeling} 
The Masked Language Modeling (MLM) pre-training objective involves recovering the original document given a corrupted one~\cite{devlin2018bert}. We choose MLM as a baseline since we can formulate code adaptation as a masked token infilling task given surrounding left and right contexts.

\textbf{Model Training}
We utilize the standard MLM objective~\cite{devlin2018bert}, in which 15\% of tokens from the original sequence is selected uniformly at random. Of the selected tokens, 80\% are replaced with \texttt{[MASK]} symbols, 10\% are left unchanged, and the remaining 10\% are substituted by random tokens from the vocabulary. In source code, this results in masking not only identifier names, but also delimiters, and programming language keywords since the token lexical types are disregarded. We train a RoBERTa model~\cite{liu2019roberta} from scratch having 12 layers in the encoder, 12 attention heads, and a hidden dimension of 768. 

\textbf{Recovering Identifier Names} During inference, the MLM for code adaptation model predicts each masked identifier name independently. Then for each pasted code snippet, all the results are summarized into a dictionary mapping masks to recovered identifier names.

\subsection{DOBF Model}
Lachaux~\emph{et al.}~\cite{lachaux2021dobf} showed that self-supervised pre-training objective for programming languages based on deobfuscation (DOBF) achieves superior results on multiple tasks such as code search, code summarization, and unsupervised code translation, as compared to the natural language specific objectives like MLM~\cite{devlin2018bert, liu2019roberta} or denoising pretraining~\cite{lewis2019bart}. We utilize DOBF as a baseline for code adaptation task.

\textbf{Model Training} 
DOBF leverages syntactic structure of programming languages, which is achieved by obfuscating code snippets via replacing class, function, and variable names with special symbols: \texttt{CLASS\_NN}, \texttt{FUNC\_NN}, and \texttt{VAR\_NN}, in which \texttt{NN} is a unique number. The DOBF model training objective is to recover the original code snippet given an obfuscated version with the special mask symbols. We pretrain a transformer model having 6 encoder and 6 decoder layers, and 8 attention heads for both encoder and decoder. 

\textbf{Identifier Deobfuscation} At inference time, the model predicts names for obfuscated class, function, and variables in pasted code. The model output is serialized as a sequence of tokens where entries for each mask symbol are separated by a delimiter ``|". The resulting string is postprocessed into a dictionary of masks mapped to identifier names.

\subsection{CodeT5 Model}

CodeT5~\cite{wang2021codet5} is a unified pre-trained encoder-decoder model for code related tasks. CodeT5 has been shown to outperform prior methods on program understanding tasks such as code defect detection, clone detection, and translation tasks between source code and natural language summaries. We use the CodeT5-small version of the model as our baseline since it has a similar configuration with 6 encoder layers and 6 decoder layers.

\textbf{Identifier names re-masking and generation} Before inference, each anonymized identifier names is re-masked as a special pre-defined token $\langle extra\_id\_0\rangle$. Then CodeT5 model predicts each re-masked identifier name independently. The results for each pasted code snippet are summarized into a dictionary that maps the original masks to generated identifier names.

\section{\thistool{} Model}
\label{sec:model}

We now introduce \thistool{} --- a dedicated transformer-based model for the code adaptation task. Given a partially written preexisting context and a code snippet copied from an external source, it tries to re-wire the pasted snippet to form a syntactically and semantically correct program. The key technical novelty of \thistool{} is twofold: (1) the model is trained with a specialized dataflow-aware deobfuscation objective, and (2) it introduces a parallel-decoder transformer architecture, a model variant that decodes masked symbols independently, allowing for selective code snippet adaptation. In addition to using a traditional transformer encoder-decoder model, which we refer to as \thistool{} uni-decoder (AP-uni) model,  the \thistool{} parallel-decoder (AP-parallel) variant allows to encode context snippets once but decode a target mapping for each adapted variable independently, leveraging ``define-use'' relationships across masked symbols. 

\begin{figure*}[tb]
 \includegraphics[width=0.85\textwidth]{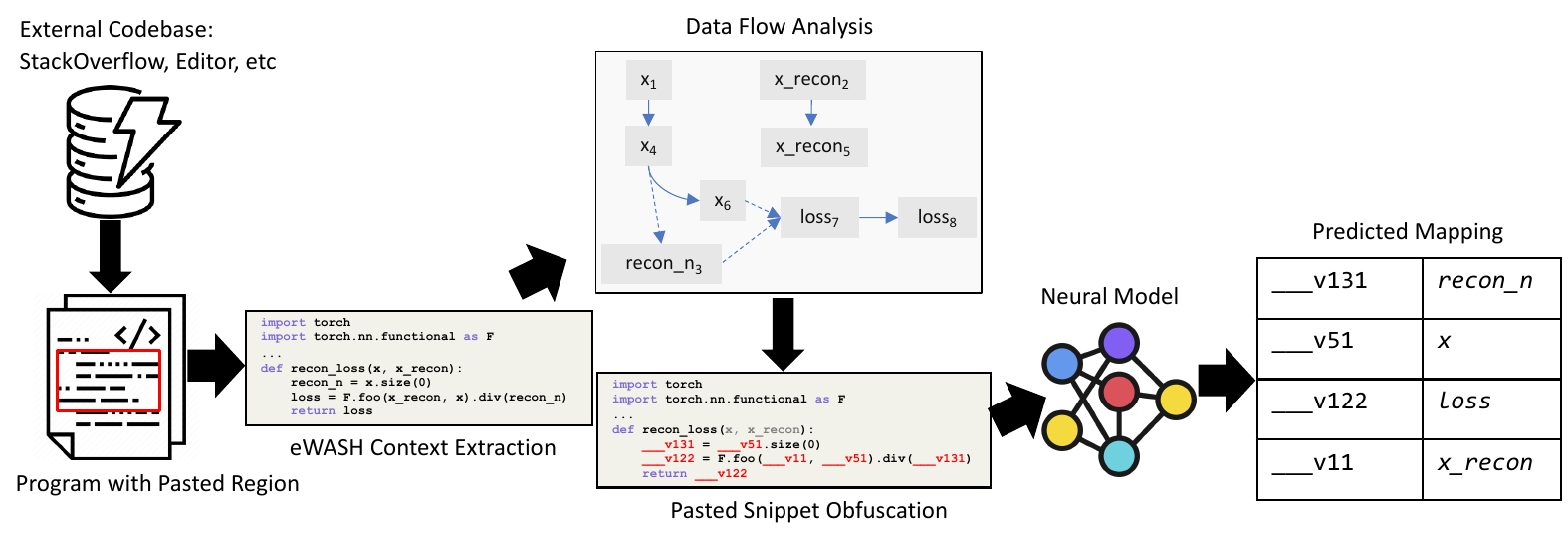}
 \caption{An overview of \thistool{} architecture: given a program with a pasted region, we first extract and prioritize syntax hierarchies most relevant for the learning task, perform data-flow analysis, and finally anonymize the pasted region. The resulting program serves as input for neural model. The output of \thistool{} is serialized as a sequence of tokens with entries separated by a delimiter symbol ``|''}
 \label{fig:architecture}
\end{figure*}

\subsection{\thistool{} Training} 
\label{sec:deobfuscation-pre-train}
Since there is no large (supervised) dataset that matches the task, we propose a self-supervised training objective.
Given some preexisting code, (e.g. a file from an open-source project) we randomly pick one or more contiguous statements as the pasted snippet, and anonymize all occurrences of a variable within the snippet to an uninformative name of the form \texttt{\_\_\_vNN}, where \texttt{NN} is a unique number. We randomize the order of masks, to prevent the model from memorizing the order and numerics. Note that this anonymization preserves the data-flow relationships \emph{within} the pasted snippet. The surrounding code becomes the context and remains unchanged. 
Finally, we train the model to match the anonymized variables to those defined in the original code. 
This self-supervised setting allows us to create a large training dataset required by modern transformers.

Variables in pasted code can belong to the bound or free categories. Free variables could arguably be named after an arbitrary non-conflicting name without changing functionality of a program. However, \thistool{} aims to predict a descriptive, useful variable name following best coding practices. \autoref{fig:architecture} illustrates the learning task: a pasted code snippet is anonymized by renaming the variables declared in the context block (\textit{x}, \texttt{x\_recon}) and free variables (\texttt{recon\_n}, \texttt{loss}) with symbols \texttt{\_\_\_v11}, \texttt{\_\_\_v51}, \texttt{\_\_\_v122}, \texttt{\_\_\_v131}. 

We assign mask symbols to variable identifiers at a granularity of whole code tokens, which in practice may map to multiple BPE subtokens. Differently from DOBF~\cite{lachaux2021dobf}, the surrounding preexisting code context is \emph{not} anonymized, which allows the model to attend to the existing identifier names that are defined in scope. Indeed, our analysis shows that 77.1\% of the anonymized variables are bound variables. Besides that, we do not encode lexical type information into masks, which is different from DOBF. An important distinction of the \thistool{} and DOBF pretraining objectives from the standard MLM is for each mask symbol to be allowed to appear multiple times throughout the input sequence.

\subsection{Neural Model Architecture}

\begin{figure}[t]
    \centering 
        \includegraphics[width=0.85\columnwidth]{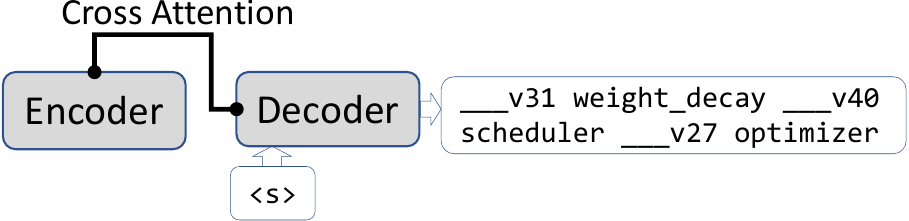}
        \caption{Uni-Decoder: A traditional encoder-decoder transformer generates a single sequence with the predicted names of the anonymized variables.}
        \label{fig:unidecoder}
\end{figure}

\begin{figure}[t]
    \centering 
        \includegraphics[width=0.75\columnwidth]{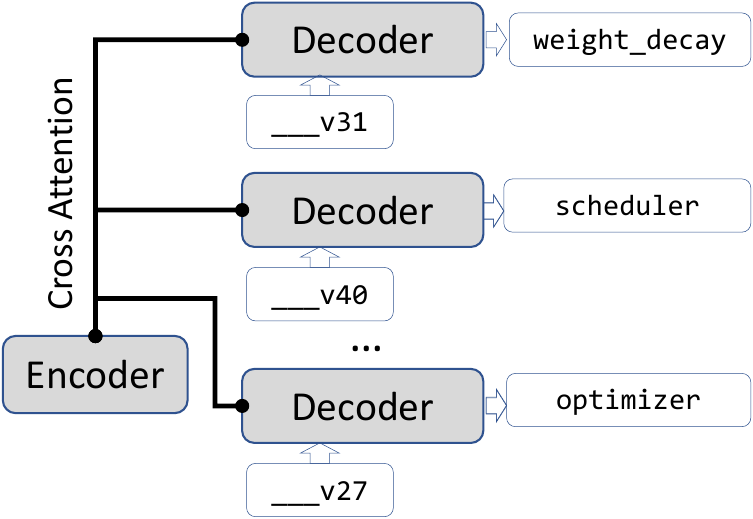}
       \caption{Parallel-decoder: Multiple weight-tied decoders are instantiated and each accepts as initial input the target variable to predict.}
       \label{fig:parallel-decoder}
\end{figure}

\paragraph{Traditional (Uni-decoder) Transformer}
\label{sec:unidecoder}
Traditional transformer~\cite{vaswani2017attention} has a single decoder composed of multiple layers of multi-headed self-attention and cross-attention blocks followed by a two-layer multi-layer perceptron (MLP), as shown in \autoref{fig:unidecoder}. In a machine translation setting, an input sequence is encoded by the encoder and is attended to by the cross-attention block of the decoder. 

The uni-decoder model follows a standard autoregressive decoder formulation. Given an input sequence composed of the context snippet and an anonymized pasted snippet, the task for the model is to learn an injective mapping $\mathcal{A}: B\rightarrow V\cup \Sigma$ that maps each variable in the pasted snippet to a unique symbol in the context or declares a new variable not currently defined. The learnt mapping, $y = \mathcal{A}(x)$, is represented as a string of the following format: $"\texttt{\_\_\_v31}~ \texttt{weight\_decay}~ |~ \texttt{\_\_\_v40}~ \texttt{scheduler}~ |$ $\texttt{\_\_\_v27}~ \texttt{optimizer}"$. The uni-decoder models the following probability distribution:
\begin{equation}
    P(y_i; \mathcal{E}(x), y_{i-1}), i = 1 ... N,
\end{equation}
where $N$ is the number of symbols to de-anonymize.

\paragraph{Parallel-decoder Transformer}
\label{sec:multidecoder}
While the uni-decoder model allows us to de-anonymize all variables jointly, it does not calculate an individual probability for each symbol, i.e., in the uni-decoder model the probability of the name of each symbol is conditioned on the ones predicted before. Thus we cannot estimate the uncertainty of de-anonymizing any particular symbol. For example in \autoref{fig:architecture}, de-anonymizing might be straightforward for all symbols except from \texttt{\_\_\_v122}. However, the uni-decoder model would assign a low probability to the entire output sequence.
To alleviate this problem, we propose a parallel-decoder model -- a variant of the parallel-decoder transformer~\cite{guo2022learning} -- which creates a copy of the decoder (with shared weights) for each anonymized symbol in the anonymized pasted snippet. Then, each decoder predicts a name independently from the rest, factorizing the output distribution per-symbol. This transformer formulation allows to perform code snippet adaptation \textit{selectively} -- surfacing only model predictions having probabilities above a certain threshold and outputting "holes" in places where the model is uncertain. See RQ3 \autoref{sec:thresholds} for more details on selective code adaptation and thresholding.

Specifically, as shown in \autoref{fig:parallel-decoder}, the parallel-decoder consists of a standard encoder $\mathcal{E}$ encoding the input sequence $x$ once (as in the uni-decoder model). Then, for each sample we spawn $N$ decoders (where $N$ is the number of distinct anonymized symbols in an input $x$) tasked with de-anonymizing each symbol independently. In lieu of the standard start-of-sequence token, each decoder accepts as initial input the name of the anonymized symbol that it should predict. This factorizes the output probability distribution as
\begin{equation}
    \prod_{i = 1}^{N} P(y_i; \mathcal{E}(x), var_{i}),
\end{equation}
where $N$ is the number of symbols being de-anonymized.
Finally, note that due to the quadratic compute and memory performance the parallel-decoder model has a reduced memory footprint compared to the uni decoder ($\mathcal{O}(nk^2)$ vs $\mathcal{O}(n^2k^2)$ where $n$ is the number of anonymized variables and $k$ is the size of the longest sequence of subtokens), since the predicted sequences are shorter and there is no attention among them. These advantages come at the cost that each anonymized variable is predicted independently from the others. See implementation details and computing infrastructure used in \autoref{sec:implementation}.

\subsection{Context Prioritization} 

A source code file may have nested scopes and references to other external libraries or other files, and to accurately adapt a pasted code snippet a model must leverage knowledge across different parts of the file. The length of source code files will often exceed the fixed-length window of transformer models, which could potentially lead to a loss of information relevant for learning to adapt. To overcome this limitation we utilized eWASH~\cite{clement2021long} to prioritize syntax hierarchies which are most relevant to the pasted snippet region. Extracting syntactic hierarchies from the entire source code files, as opposed to the tokens immediately preceding the pasted snippet location, we are able to retain bound variables, such as class-level fields and method arguments, which are highly relevant for de-anonymizing variables in the pasted code snippet. Starting with a concrete syntax tree of a source file, we organize and prioritize class-level and method-level syntactic elements such as global import statements and assigned values, class attributes, method signatures, class docstring, and global expressions in the input. A special symbol ``\texttt{$...$}" is introduced to denote locations in the code where syntax hierarchies are truncated. 

\subsection{Implementation Details}
\label{sec:implementation}

Both the uni-decoder and parallel-decoder \thistool{} transformer models have 12 layers, including 6 encoder transformer blocks and 6 decoder transformer blocks. Each encoder/decoder block has 6 attention heads. The vocabulary size of BPE tokenizer is 50000. We utilize the HuggingFace's byte-level BPE tokenizer\footnote{\url{https://huggingface.co/docs/transformers/tokenizer\_summary\#bytelevel-bpe}}. The dimension of embedding space is 512. We train \thistool{} using the Adam stochastic optimization scheme with the zero redundancy optimizations~\cite{rajbhandari2020zero}. The base learning rate is $5\times 10^{-5}$ and cumulative batch size is 288.  



The offline training module of \thistool{} is implemented in Python, which makes use of CUDA 11.4, GPU accelerated computing primitives from CuDNN 8.2 library, Pytorch 1.10.0, and the NVIDIA collective communication library (NCCL). The model is trained on a Kubernetes cluster of 6 worker nodes, each having 8 V100 GPUs with 32 GB device memory.

\section{Evaluation}

\paragraph{Metrics}
To assess code adaptation performance, we identify following evaluation metrics: (i) exact string match of an adapted code snippet prediction as compared to the ground truth user code, and (ii) \textit{ValidPaste}: 
a measure of semantic validity of paste instances. We define exact match accuracy as a ratio of the number of code snippets de-anonymized correctly divided by the total: $\texttt{N}_{\texttt{correct}}/\texttt{N}_{\texttt{total}}$. When studying selective code adaptation in presence of a threshold we consider exact match precision and recall. The ValidPaste accuracy is calculated as a ratio of predicted code snippets with all bound variables de-anonymized correctly to the total: $\texttt{N}_{\texttt{correct\_bound\_variable}}/\texttt{N}_{\texttt{total}}$. Among the two metrics, the exact match accuracy is stricter, as it measures both the correctness of identifier adaptations which are defined in the preexisting context snippet as well as the exact naming of new variables declared in the pasted snippet. Fortunately, the latter does not affect the semantics correctness of source code, because arbitrarily renaming the variables declared and used within a pasted snippet range would keep a paste instance semantically equivalent.

In addition to the above evaluation metrics which are calculated per pasted code snippet, we calculate per-variable accuracy of de-anonymization -- denoted as \textit{VarAcc}, and the accuracy of predicting the variable category (bound or free)  -- \textit{CatAcc}.

Finally, given that each anonymization symbol $\texttt{\_\_\_vNN}$ in a pasted snippet may correspond to a code token composed of multiple subtokens differing by the casing conventions only (i.e. \texttt{camel}-\texttt{Case} versus \texttt{snake\_case}), we consider a casing-insensitive evaluation metrics. Namely, average harmonic mean of the precision\footnote{Percentage of predicted free variable subtokens are correct with respect to the ground truth.} and recall\footnote{Percentage of actual free variable subtokens are predicted as correct.} for retrieving the original case-insensitive free variable subtokens (Avg. F1-Subtoken). Free variable subtokens are determined by breaking the corresponding free variable name into subtokens using uppercase letters for \texttt{camelCase} and underscores for \texttt{snake\_case}.
\begin{table}[t]
    \centering
    \small
    \begin{tabular}{lll}\toprule
        & \textbf{Accuracy} & \textbf{ValidPaste}
         \\\midrule 
        MLM-Adapt & 22.5  &  29.7  \\ 
        DOBF & 28.0 &  32.4 \\     
        CodeT5 & 39.5 & 48.1 \\
        \\\midrule 
        AP-uni & \textbf{67.8} &  \textbf{83.9}  \\ 
        AP-parallel & 61.2 & 78.0  \\
    \bottomrule
    \end{tabular}
    \caption{ \thistool{} and baseline models evaluated in terms of top-1 accuracy of code snippet adaptation and the ValidPaste metric.
   \label{tab:topk}}
\end{table}

\begin{table}[t]
    \centering
    \small
    \begin{tabular}{lll}\toprule
         & \textbf{AP-parallel} & \textbf{AP-uni} \\
         \cmidrule{2-3}

        \multicolumn{3}{l}{\textbf{Bound Variables}} \\\midrule
        VarAcc (all | previous var incorrect) & 87.4 | \textbf{73.7} & \textbf{89.9} | 70.5  \\
        \midrule
        CatAcc (all | previous var incorrect) & 94.2 | 89.8 & \textbf{96.7} | \textbf{91.7} \\
        \\
        \multicolumn{2}{l}{\textbf{Free Variables}} \\\midrule
        VarAcc (all | previous var incorrect) & 40.1 | \textbf{25.0} & \textbf{45.4} | 23.3  \\
        \midrule
        CatAcc (all | previous var incorrect) & 91.0 | \textbf{87.6} & \textbf{91.5} | 87.3 \\ 
        \midrule
        Avg. F1-Subtoken (all | previous var incorrect) & 49.2 | \textbf{35.7} & \textbf{51.6} | 31.7 \\
    \bottomrule
    \end{tabular}
    \caption{Evaluation of \thistool{} model variants in terms of per-variable top-1 accuracy of de-anonymization (VarAcc) and the top-1 accuracy of predicting the variable category (CatAcc) in general cases and previous variable incorrectly adapted cases. We perform two accuracy measurements: for all variables in each adapted code snippet, and only for variables which appear after an incorrectly predicted variable in a snippet.}
    \label{tab:eval-results}
\end{table}


\subsection{Results}
In this section we discuss the results for the 3 research questions.

\textbf{RQ1. How effective is \thistool{} in adapting the copy-pasted/ported code snippets?}

As seen in \autoref{tab:topk}, both \thistool{} model variants beat baselines across all the evaluation metrics. The best-performing \thistool{} variant, the uni-decoder model (AP-uni), achieves 67.8\% exact match accuracy of code adaptation, surpassing the top performing baseline (CodeT5) by 28.3\%. Furthermore, it exceeds CodeT5 results in terms of the ValidPaste accuracy metric, which measures the percentage of semantically valid paste instances, by 35.8\% at 83.9\%. The advantages of AP-parallel model variant are explained in RQ2 and RQ3 sections.

\textbf{RQ2. How does \thistool{} perform on bound variables and free variables?}

Correctly adapting a code snippet requires to de-anonymize several variables, which may be \textit{free} or \textit{bound}. To disentangle the performance differences among AP-uni and AP-parallel we turn to per-variable evaluation. In addition, we factor out a scenario in which \thistool{} makes an incorrect de-anonymization for a variable at a preceding span in a code snippet, with the purpose of assessing if the model can recover from previous mistakes, and provide accurate adaptations for the remainder of the code snippet.

As seen in \autoref{tab:eval-results}, AP-uni is more accurate in predicting variables names for each variable category, showing 2.5\% higher VarAcc for bound and 5.3\% higher VarAcc for free variables as compared parallel-decoder model variant, and yielding 2.4\% higher average F1-Subtoken metric, which is our variable casing-insensitive metric. This observation persists for the variable category prediction accuracy, as the AP-uni achieves better performance for both bound and free variables.

However, for the scenario in which incorrect variable adaptations have been made, AP-parallel shows advantages across almost all measured metrics by achieving 3.2\% higher VarAcc of predicting the next bound and 1.7\% higher VarAcc and 4\% higher average F1-Subtoken metric of predicting next free variables. For the next variable category prediction accuracy, the AP-parallel shows a slightly better performance for the free variable case, but not for bound variables.

\textbf{RQ3. Can we improve the precision of \thistool{} by introducing a confidence threshold on its predictions?}
\label{sec:thresholds}

We perform a systematic study of the \thistool{} uni-decoder and parallel-decoder model predictions as a function of the confidence threshold to determine if it can improve code adaptation quality, resulting in higher precision and semantic validity metrics. We expect low confidence predictions to be of inferior quality or to violate program semantics. To estimate its potential effect, we have measured the model performance in 20 equal-sized confidence threshold bins. We evaluate \thistool{} in terms of the exact match precision and the ValidPaste precision metrics as a function of the confidence threshold on the model predictions. To facilitate this study, we exponentiated the log-probability scores of the output sequences generated by \thistool{} to fit into the 0 to 1 probability range. As introduced in~\ref{sec:model}, the parallel-decoder model creates a copy of the decoder for each anonymized symbol in a pasted snippet then each decoder predicts a name independently from the rest. As such, each variable prediction receives its own prediction probability. On the other hand, the uni-decoder model variant de-anonymizes all variables jointly, assigning a single probability to all the predicted variables in a given pasted snippet. Our analysis shows that 89\% of the predicted variables have probabilities in the 0.7---0.985 range. 

\begin{figure}
     \centering
     \begin{subfigure}[b]{0.95\columnwidth}
         \centering
         \includegraphics[width=0.95\columnwidth]{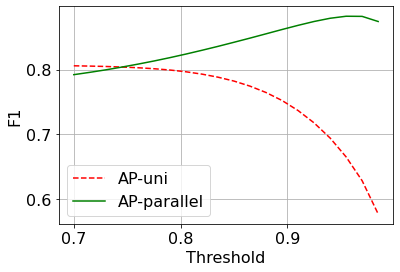}
         \caption{F1 score of exact match as a function of threshold.}
         \label{fig:em-vs-threshold}
     \end{subfigure}
     \hfill
     \begin{subfigure}[b]{0.95\columnwidth}
        \includegraphics[width=0.95\columnwidth]{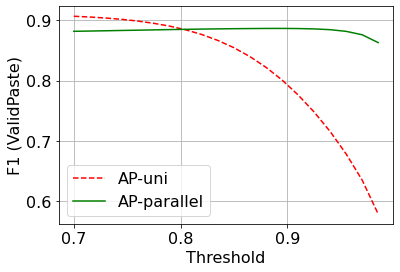}
        \caption{F1 score of ValidPaste as a function of recall.}
        \label{fig:pvp-vs-threshold}
     \end{subfigure}
     
     \caption{
        Comparison of the \thistool{} parallel-decoder (AP-parallel) and uni-decoder (AP-uni) model variants in terms of the F1 scores calculated as harmonic mean of the exact match precision and recall, as well as the ValidPaste precision and recall. The recall metric is computed as a fraction of the variable predictions passing the confidence threshold. The confidence threshold is varied in the $[0.7, 0.985]$ range. 
     }
     \label{fig:empvp-vs-threshold}
\end{figure}

\autoref{fig:em-vs-threshold} shows F1 score, calculated as a harmonic mean between the exact string match precision and recall, as a function of the confidence threshold for uni-decoder and parallel-decoder variants of the model. The recall is defined as a fraction of the variable predictions passing the confidence threshold, which is varied in the 0.7---0.985 range. As seen, increasing probability threshold improves the de-anonymization performance for the AP-parallel model, achieving its best F1 score. However, for AP-uni, increasing probability threshold results in a worse de-anonymization performance and a much lower F1 score. This observation is expected as AP-parallel predicts each variable individually with a separate log-probability score, which makes it easier to filter out bad adaptation suggestions without compromising the good ones in the same given pasted snippet. Similarly, \autoref{fig:pvp-vs-threshold} shows the F1 score calculated as a harmonic between ValidPaste precision and recall as a function of confidence thresholds in 0.7---0.985 range. In the case of semantic validity of the paste instances, we observe the same trend: introducing a confidence threshold leads to an improvement of F1 scores for AP-parallel, but not for AP-uni. As seen, the F1 score is monotonously increasing with threshold for AP-parallel but decreasing with threshold for AP-uni. This motivates the selective code adaptation, which is a particularly useful practical deployment scenario where a developer assistant tool may sacrifice recall for precision by performing selective code snippet adaptation, surfacing only model predictions with probabilities above a certain threshold and outputting "holes" in places where the model is uncertain. 

\begin{figure}
     \centering

     \begin{subfigure}[b]{0.95\columnwidth}
         \centering
         \includegraphics[width=0.95\columnwidth]{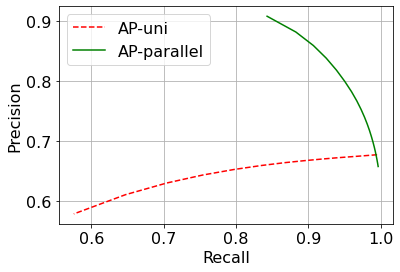}
         \caption{Precision vs. recall as a function of threshold.}
         \label{fig:em-vs-threshold-2}
     \end{subfigure}
     \hfill
     \begin{subfigure}[b]{0.95\columnwidth}
        \includegraphics[width=0.95\columnwidth]{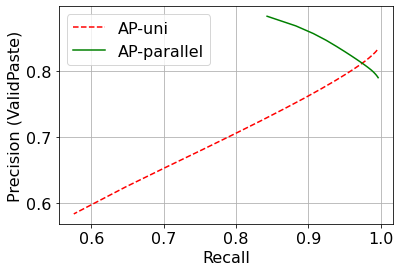}
        \caption{ValidPaste precision vs. recall as a function of threshold.}
        \label{fig:pvp-vs-threshold-2}
     \end{subfigure}
     
     \caption{
        Comparison of the \thistool{} parallel-decoder (AP-parallel) and uni-decoder (AP-uni) model variants in terms of the exact match accuracy (Accuracy) and percentage of valid paste instances (ValidPaste) as a function of recall, which is computed as the percentage of the variable predictions passing the confidence threshold. The confidence threshold is varied in the $[0.7, 0.985]$ range.
     }
     \label{fig:empvp-vs-threshold-2}
\end{figure}

It is remarkable to note that our parallel-decoder model variant (AP-parallel) achieves a superior performance over traditional transformer-based AP-uni when predictions are filtered according to a confidence threshold, which provides a potentially powerful knob to tune the model performance for the selective code adaptation scenario in developer tools. As it shows in \autoref{fig:em-vs-threshold-2}, when confidence threshold is 0.96, precision is significantly improved to 85.9\% with 90\% of the variable predictions covered, while precision of AP-uni is significantly decreased to 66.7\%. This trend persists in \autoref{fig:pvp-vs-threshold-2}, which shows that with 90\% of the variable predictions covered, AP-parallel achieves 85.7\% ValidPaste precision, while AP-uni achieves only 76.0\%. As a summary, AP-parallel adopting a confidence threshold yields the best performance by improving the original precision and ValidPaste achieved by AP-uni by 8.1\% and 1.8\%, respectively. 

\section{User Evaluation}

\subsection{User Study Design}

To demonstrate the value of \thistool{} in practice, we perform a user study among software developers of varying level of programming experience, asking them to manually port code snippets between real-world software projects. For comparison, we adapt the same set of code snippets with \thistool{}. A code snippet is considered to be adapted correctly by the study participants only if all the identifier names are reviewed and adapted correctly.

To mine candidate code snippets for the user study, we select following five open-source Python projects: RethinkDB, UnityPack, CastXML, MultiReql, VINet. These projects are being actively developed and happen to contain several code clones, which we identified by the SourcererCC\cite{SourcererCC} tool as exact matches or structurally similar code snippets to either code from Stack Overflow posts or open-source repositories. For each selected code clone, we anonymize all the identifier names with special symbols for the purpose of the user study. To ensure the quality of candidate code clones, we introduce the following selection criteria:
\begin{itemize}
\item Code clones should have passed a code review, which means the code clones with adapted identifier names must have been reviewed in a Pull Request and merged into the main branch after approval issued by a reviewer.

\item Code clones should have at most five adaptable variables, with bound variables comprising at least a half of the variables. Whether a variable is bound or free can only be determined by understanding a given surrounding code context. We conjectured, that having more variables to adapt may be too complex to finish within the interview time slot.
\end{itemize}

To create a pool of study participant candidates, we identified and recruited 22 Python developers, who are professional software engineers with 2--25 years of working experience and never used \thistool{} before. Finally, 20 out of the 22 contacted developers agreed to participate in this study. To evaluate code adaptation in diverse situations, each code adaptation example is classified as either ``easy" or ``hard". An example is considered hard if (i) it has more than 2 variables to adapt, and (ii) it uses non-standard Python libraries or developer-custom APIs. In total, we have collected 100 examples for user evaluation, with 30 of which classified as ``easy" examples, and 70 ``hard" examples.

\subsubsection{User Study Interface} 

The interface for user study is designed as a Visual Studio Code extension, in which participants can browse each example, do manual adaptation and operates \thistool{}. In addition, participants can click through their results and \thistool{}’s results to evaluate adaptation accuracy and validity. They can view the original adaptations if needed. For easy access at client side, \thistool{} model is deployed in a cloud-based web server with one V100 GPU.

\subsubsection{User Study Protocol}

We conduct a real-world application study by interviewing participants one by one, remotely, through a video conferencing tool, over a 30-minute time slot. We divide the study participants into two groups: Every participant in group \#1 is asked to complete each code adaptation example by manually predicting names of anonymized identifier names given surrounding code context. One of the authors was on the video conferencing call to help participants navigate the examples, asking participants clarifying questions, and recording how long it took them to adapt each example. In the meantime, each participant from group \#2 is introduced to the \thistool{} interface and then timed as they adapt the same set of examples automatically, using the tool. Finally, we ask all participants to evaluate the accuracy by following the same evaluation metrics specified in Section VI-a. Additionally, we calculate the average time taken by study participants and \thistool to perform adaptation.


\begin{table}[t]
    \begin{subtable}{.5\textwidth}
    \centering
    \small
    \begin{tabular}{llll}\toprule
        & \textbf{Accuracy} & \textbf{ValidPaste} & \textbf{Time (seconds)}
         \\\midrule 
        AP-uni & \textbf{90.0} &  \textbf{96.3} & \textbf{1.0}  \\ 
        Study participants & 65.0 &  69.3 & 137.5 \\
    \bottomrule
    \end{tabular}
    \caption{All Snippets
   \label{tab:user-study-all}}
    \end{subtable}
    \begin{subtable}{.5\textwidth}
    \centering
    \small
    \begin{tabular}{llll}\toprule
        & \textbf{Accuracy} & \textbf{ValidPaste} & \textbf{Time (seconds)}
         \\\midrule 
        AP-uni & \textbf{100.0} &  \textbf{100.0} & \textbf{1.0}  \\ 
        Study participants & 63.8 &  64.3 & 101.2 \\
    \bottomrule
    \end{tabular}
    \caption{Snippets utilizing PyTorch library}
    \label{tab:hard-pytorch}
    \end{subtable}
    \begin{subtable}{.5\textwidth}
    \centering
    \small
    \begin{tabular}{llll}\toprule
        & \textbf{Accuracy} & \textbf{ValidPaste} & \textbf{Time (seconds)}
         \\\midrule 
        AP-uni & \textbf{83.3} &  \textbf{93.8} & \textbf{1.0}  \\ 
        Study participants & 63.3 &  70.0 & 231.3 \\
    \bottomrule
    \end{tabular}
    \caption{Snippets utilizing developer-custom APIs}\label{tab:hard-others}
    \end{subtable}
    \caption{ Code adaptation performance on all examples and examples of hard level of difficulty by \thistool{} and the study participants evaluated in terms of top-1 accuracy, ValidPaste, and time to solution.
   \label{tab:user-study-hard}}
\end{table}

\vspace{-3mm}

\subsection{User Study Results}

\textbf{RQ4. How do developers and \thistool{} perform on adapting real-world source code?}

Tab.~\ref{tab:user-study-all} illustrates a performance gap between manual code adaptation (by the study participants) and \thistool{}. Surprisingly, out of the 100 examples included in the study, participants achieved only 65\% accuracy, averaging 137.5 seconds per example, as compared to 90\% accuracy, and only a second per sample on average by \thistool{}. 
In terms of ValidPaste, \thistool{} achieved 96.3\%, surpassing manual code adaptation by the participants by 27\%. By analyzing results of the user study, we have identified a few aspects suggesting why \thistool{} suggests more accurate code as compared to human experts:

\textbf{Knowledge about special API:} 40\% (30/70) of the hard code adaptation examples use PyTorch library by invoking PyTorch API functions and instantiating variables of \textit{torch.tensor} type. 70\% (7/10) of the participants indicated that they would need to learn PyTorch library before they can perform accurate code adaptation. As Tab.~\ref{tab:hard-pytorch} shows that study participants only achieve 63.8\% in accuracy and 64.3\% in ValidPaste precision, while \thistool{} generates perfect suggestions.

\textbf{Familiarity of code context:} The quality of code adaptation also depends on how much do study participants understand the surrounding code context. As shown in Tab.~\ref{tab:hard-others}, study participants found it difficult to adapt code which uses custom methods and types. Developers who are not the direct contributors or reviewers of these custom APIs needed to spend more time to read and understand the implementation details of each custom API used in the surrounding context. 

\begin{figure}[t]
    \centering 
        \includegraphics[width=0.85\columnwidth]{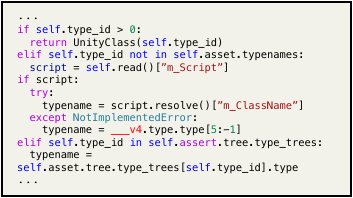}
       \caption{A code adaptation example which user study participants have failed to adapt  due to a lack of familiarity with the code context.}
       \label{fig:familiar-example}
\end{figure}

As shown in \autoref{fig:familiar-example}, an example in the study that participants failed to adapt is renaming the anonymized variable \texttt{\_\_\_v4} to \textit{script}. To make this adaptation, developers would need to learn about the return type and corresponding attributes of custom method \textit{self.read()}. \thistool{} was able to successfully adapt this example by inferring the aforementioned information from context.  

\begin{figure}[t]
    \centering 
        \includegraphics[width=0.85\columnwidth]{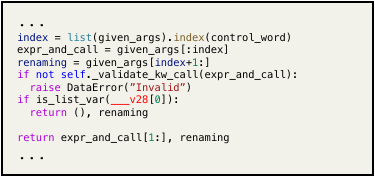}
       \caption{
       An example of incorrect prediction by \thistool}
       \label{fig:incorrect-example}
\end{figure}

\thistool{} did not generate valid suggestions for only 3.7\% of the variables. We manually inspected the incorrect predictions and identified that \thistool{} mistaken re-assigned variables for their original variable names. \autoref{fig:incorrect-example} shows an example in which \thistool{} predicted a variable \textit{\_\_\_v28} as \textit{given\_args}, instead of the correct answer \textit{expr\_and\_call} that \textit{given\_args} was re-assigned as. To reduce this kind of incorrect suggestions, \thistool{} may be re-trained taking into account the distance from candidate variables to the point of adaptation.

\vspace{-3mm}

\section{Related Work}

Our work relates to the literature on transformers for code, code deobfuscation, and program repair.

\textit{Transformers for Code.} Self-supervised pretraining objectives have shown to be highly effective for code intelligence tasks. Feng~\emph{et al.}~\cite{feng2020codebert} proposed CodeBERT, a RoBERTa-based MLM model pretrained on code, for natural language code search and code documentation generation tasks. Guo~\emph{et al.}~\cite{guo2020graphcodebert} extended this work by proposing an GraphCodeBERT to predict edges in a dataflow graph. Svyatkovskiy~\emph{et al.}~\cite{svyatkovskiy2020intellicode} introduced GPT-C, a generative transformer model trained on source code for line-level code completion. Clement~\emph{et al.}~\cite{clement2020pymt5} proposed a Python method text-to-text transfer transformer (PyMT5) for method generation and code summarization. Tufano~\emph{et al.}~\cite{tufano2020generating} applied BART~\cite{lewis2019bart} to learn to generate assert statements. Chen~\emph{et al.}~\cite{chen2021evaluating} introduced Codex, a large language model trained on all publicly available code from GitHub, to support developers in tasks such as code completion and method generation. 

\textit{Code Deobfuscation.} Informative identifier names can make code more understandable and easier to maintain~\cite{butler2009relating}. This motivated researchers to study deobfuscation of identifier names. Allamanis~\emph{et al.}~\cite{allamanis2014learning,allamanis2015suggesting} leverage n-gram model to suggest identifier names such as method and class names. Raychev~\emph{et al.}~\cite{raychev2015predicting} present a approach for predicting program properties via a probabilistic model learnt from massive codebases. Bavishi~\emph{et al.}~\cite{bavishi2018context2name} presents a recurrent neural networks based technique to infer natural identifier names for minified names. Alon~\emph{et al.}~\cite{alon2018general} train a neural networks with features extracted from abstract syntax tree paths to suggest variable and method names. David~\emph{et al.}~\cite{david2020neural} use augmented representation obtained from static analysis to predict procedure names in binary files. Recently, Lachaux~\emph{et al.}~\cite{lachaux2021dobf} introduced a deobfuscation pre-training objective for programming languages (DOBF) model, which outperforms other approach in many source code related tasks. However, none of these papers aimed to adapt variable names in pasted code. Furthermore, our evaluation results show than our proposed \thistool{} model outperforms  deobfuscation for the code adaptation task. 

\textit{Program Repair and Synthesis.} Program repair, especially variable misuse repair, is the most relevant task to code adaptation. Ray~\emph{et al.}\cite{ray2013detecting} classifies code adaptation errors into 5 following categories:  inconsistent control flow, inconsistent renaming, inconsistent data flow, redundant operations, and all others. Showing, that identifier renaming amounts to nearly a half (41--48\%) of all the porting errors, while inconsistent data flow amounts for another 14--28\%. In this paper, we are targeting both of these classes of code adaptation bugs. Pradel~\emph{et al.}~\cite{pradel2018deepbugs} propose DeepBugs to use an MLP over a limited window of code tokens to detect wrong operators, operands, and argument swappings. Dinella~\emph{et al.}~\cite{dinella2020hoppity} present Hoppity to learn to perform graph transformation to replicate code refactoring, introducing functionality, bug fixing. There are several approaches proposed to tackle variable misuse bugs. Allamanis~\emph{et al.}~\cite{allamanis2017smartpaste} designed a set of deep neural models to adapt given snippet to surrounding code. Vasic~\emph{et al.}~\cite{vasic2019neural} propose a joint approach to classify the variable locations as faulty or correct and then replace the faulty variables. Hellendoorn~\emph{et al.}~\cite{hellendoorn2019global} aim at variable misuse identification with Graph Relational Embedding Attention Transformers. Allamanis~\emph{et al.}~\cite{allamanis2021self} present BugLab to co-train a detector to detect as well as repair bugs in code and a selector to create buggy code for the detector to use as training data. Even though \thistool{} tackles a task similar to variable misuse repair, it leverages sequence-to-sequence with parallel-decoder transformer pre-training to learn programming language semantics, including the long-range dependencies of variables usages, to adapt multiple variables in pasted snippet of code.

Our work is remotely related to work on program synthesis using sketches \cite{solar2008program} and automated code transplantation \cite{barr2015automated}. However, both approaches require specifications (e.g. input-output
examples, test suites) and cannot be used to adapt code statically. These
approaches can be thought as complementary to \thistool, since it learns to statically adapt the code.

\vspace{-3mm}

\section{Deployment}

\thistool{} has been deployed initially as an extension for VSCode, and later more broadly as a feature on Visual Studio integrated development environment. Shipping a tool that transforms how software developers copy and paste code -- an action so fundamental to our daily interactions with IDE -- into a product represented its own set of challenges and benefits which we will discuss in this section.

\vspace{-3mm}

\subsection{Iterating on UX Design}

Our initial approach to surfacing the model predictions in IDE was in line with our unofficial slogan of having code that ``just runs'' after being copy-pasted from a different source. Our goal was to avoid imposing additional editing steps or cognitive actions on users.
However, our greatest strengths had quickly become our greatest weakness when \thistool{} was presented via a silent user experience: under the hood, \thistool{} was performing a significant amount of code editing and refactoring. Early testers found this silent experience ``jarring''. For one tester, a rare model blunder caused a compiler error which went unnoticed. 

\vspace{-3mm}

\subsection{Compiler Driven Adaptation}
This feedback had taught us important lessons. First, users wanted to be aware of any changes that were to happen to their code. Second, they did enjoy the changes \thistool{} made. In most cases it made their code an executable state, but they would only be satisfied after verifying the changes. Third, less edits were generally better accepted by developers as it took less cognition to verify. These findings combined dictated a new experience that would prompt users to review and accept changes required to make the code executable. The powerful C\# compiler, Roslyn, seemed a perfect companion to help achieve this experience.

\begin{figure}[t]
\centering
 \includegraphics[width=\columnwidth]{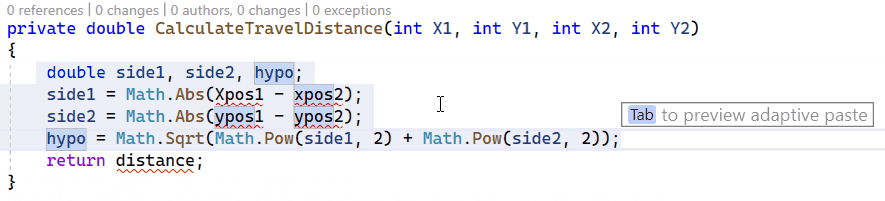}
 \caption{Show to users that suggestion is ready}
 \label{fig:vs_suggestion_ready}
\end{figure}

\begin{figure}[t]
\centering
 \includegraphics[width=\columnwidth]{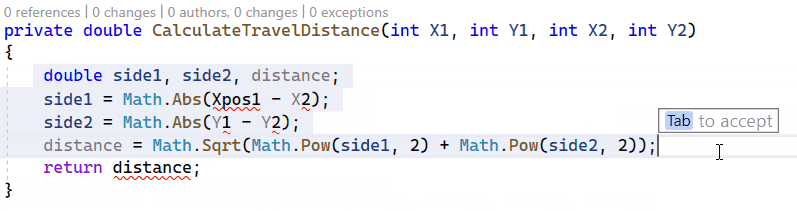}
 \caption{Inline gray text view for suggestions}
 \label{fig:vs_suggestion_preview}
\end{figure}

In Visual Studio, the C\# compiler Roslyn works as a language server to constantly keep a check on your code to find syntactic errors.
Leveraging this system, Roslyn helped find the minimal required changes for the code to at least compile successfully. By iterating through all of the suggestions generated via \thistool{}, we could show only the changes that lowered the number of compiler errors (Fig.~\ref{fig:vs_suggestion_ready}). With the users' command, an inline view would show the changes, allowing to quickly verify what the minimal changes were to improve compiler errors (Fig.~\ref{fig:vs_suggestion_preview}).

\vspace{-3mm}

\subsection{Model Deployment}
Deploying \thistool{} in VSCode and Visual Studio required careful consideration, balancing several factors including performance, privacy, scalability, and maintainability. We have selected to deploy \thistool{} on the client side leveraging ONNX runtime for local CPU execution.

Client-side deployment in IDE provides several attractive benefits. The first and foremost being the privacy. Under this scheme, the user source code never leaves the local computer. Local containment of the runtime environment removes the necessity for secure data transmission, or secure data processing on a server side. The second advantage is a lack of network latency.

To ensure that the model can be conveniently hosted in both Visual Studio and VSCode client, the underlying transformer model has been optimized by less-efficient operator replacement and INT8 format quantization. This approach has led to a reduced model size of 74.2\% (from 440MB to 114MB). Moreover, our model was measured to be fast enough to fit into a 110ms window over 90\% of the time on the laptop CPU of a Surface Book. 

\vspace{-3mm}

\section{Conclusions and Future Work}
We introduced \thistool, a transformer-based code adaptation model, which was pretrained with a specialized dataflow-aware anonymization objective. We demonstrated its ability to learn to adapt source code by capturing meaningful representations of variable usage patterns and adapting pasted snippets to the surrounding, preexisting code context. Both uni-decoder and parallel-decoder variants of \thistool{} have been shown to achieve the state-of-the-art performance in code adaptation, outperforming MLM and DOBF-based models, as well as CodeT5. We have introduced a dedicated parallel-decoder transformer model architecture -- AP-parallel -- which de-anonymizes symbols independently. We have shown that it is particularly useful in the selective code adaptation setting. Furthermore, we have shown the practical value of \thistool{} in reducing the dwell time spent on adapting source code, and helping to avoid bugs. 

Although this work had focused on code adaptation errors due to inconsistent identifier renaming or inconsistent dataflow -- the prevalent classes of code adaptation errors -- extending the scope of \thistool{} to other classes of code adaptation errors, those which insert or delete code tokens, seems possible and will be explored in the future work.

\vspace{-3mm}

\section*{Limitations}

Self-supervised data generation might not fully reflect all real-life use case scenarios of code copy-pasting studied in this work. This limitation, however, does not affect our evaluation and comparison to baselines, since we are using the same train and test split and a standard procedure to extract snippets. 

Threats to internal validity can occur in our hyperparameter configuration. Due to substantial training time requirement, we didn't perform a hyperparameter search. To address this concern, we reuse configurations suggested in the literature. In addition, we experiment with multiple \thistool{} variants with different model configurations and report their results.




\balance

\bibliography{refs}
\bibliographystyle{ACM-Reference-Format}

\end{document}